\documentclass[12pt]{article}
\title{Note on Quantum Newtonian Cosmology}
\author{Juan~M~Romero${}^{}$\footnote{e-mail: sanpedro@nucleares.unam.mx}\,
and Adolfo~Zamora${}^{}$\footnote{e-mail: zamora@nucleares.unam.mx}\\
\vspace{.51cm}\\
{\it Instituto de Ciencias Nucleares}\\
{\it Universidad Nacional Aut\'onoma de M\'exico}\\
{\it Apartado Postal 70-543, M\'exico DF 04510, M\'exico}
}
\date{\today}
\begin{document}
\pagestyle{plain}
\maketitle

\begin{abstract}
It is well known that, for pressureless matter, Newtonian and
relativistic cosmologies are equivalent. We show that this 
equivalence breaks down in the quantum level. In addition, we 
find some cases for which quantum Newtonian cosmology can be related 
to quantum cosmology in (2+1) dimensions. Two exact solutions for 
the wave function of the Newtonian universe are also obtained.
\end{abstract}


\newpage

\section{Introduction}
An interesting outcome from Newtonian cosmology is that Friedmann 
equation for pressureless matter can be obtained from it. That is,
for this kind of matter the dynamics obtained from both Newtonian 
and relativistic cosmologies is equivalent. The first work on this
topic dates back to 1934 when E~A~Milne \cite{milne:gnus} showed
that by combining fluid equations and the cosmological principle 
one gets to Friedmann equation for pressureless matter. Extensions
of this work have been carried out \cite{esos:gnus} and interestingly
it has been shown \cite{weinberg:gnus} that the same Friedmann equation 
can also be obtained from classical mechanics. A recent discussion 
on this approach can be found in Ref. \cite{myung:gnus}.

Because of the equivalence between Newtonian and relativistic
cosmologies just mentioned, it is natural to expect them also
to be equivalent in the quantum level. That is, for pressureless
matter, one would expect the wave equation of the Newtonian 
universe to coincide with that of its relativistic counterpart.
In this work we show that such a thing does not happen. In fact,
despite both cosmologies yield the same equations of motion for 
pressureless matter, they have different phase spaces. Therefore,
after implementing quantization rules, they yield different
quantum systems. By considering a matterless space, but with 
cosmological constant, we also find that Newtonian quantum cosmology
is close to quantum cosmology in a (2+1) dimensional space. To
end up, we present two cases where the wave function of the
Newtonian universe can be calculated exactly. 

The work in this paper is organized as follows: In Section 2 we
quickly review the Newtonian cosmology and write down the wave
equation of the Newtonian universe. In Section 3 we find the
differences between Newtonian quantum cosmology and quantum
cosmology in (3+1) dimensions. Some cases for which Newtonian 
quantum cosmology can be mapped to quantum cosmology in (2+1)
dimensions are found in Section 4. Section 5 presents a couple
of exact solutions for the wave equation of the Newtonian
universe and finally in Section 6 conclusions are drawn.

\section{Newtonian Cosmology}
We concentrate first in obtaining Friedmann equation from
classical mechanics. For this, let us assume we have a system
of particles interacting gravitationally. Its energy is 
therefore
\begin{equation}
E=\frac{1}{2}\sum_{i}^{n} m_i\dot r_{i}^2
-G\sum_{i>j}^{n}\frac{m_im_j}{|r_{i}-r_{j}|}
-\frac{\Lambda}{6}\sum_{i}^{n} r_{i}^2.
\end{equation} 
Now, by considering the cosmological principle we find that
$r_{i}(t)=S(t)r_{i}(t_{0})$, and thus the energy can be
rewritten as
\begin{equation}
E=\frac{1}{2}A \dot S^2
-G \frac{B}{S}-\frac{\Lambda}{6}A S^2, \label{eq:wen}
\end{equation} 
with $A=\sum_{i}^{n} m_i r_{i}^2(t_{0})$ and
$B=\sum_{i>j}^{n}\frac{m_im_j}{|r_{i}(t_{0})-r_{j}(t_{0})|}$.
It is not difficult to see that Eq. (\ref{eq:wen}) is an 
alternative form of Friedmann equation. By introducing 
the rescaling $a=\mu S$, with $\mu={\rm const}$, this equation
becomes
\begin{equation}
H^{2}=\left( \frac{\dot a}{a}\right)^{2}=-\frac{k}{a^{2}}+
\frac{\Lambda}{3}
+\frac{8\pi G}{3} \rho, \label{eq:fn}
\end{equation}
with
\begin{equation}
k=-\frac{2E\mu^{2}}{A},\quad \rho=\frac{\rho_{0}}{a^{3}},
\quad{\rm and}\quad \rho_{0}=\frac{3B\mu^{3}}{4A\pi}.
\end{equation}
Notice that if $E=0$, then $k=0$. However, if $E\not=0$ one
may take $\mu^{2}=A/2|E|$; which implies either $k=+1$ or $k=-1$.
We see then that $\mu$ can be chosen in such a way that $k$ only 
takes values $1,-1,0$; and therefore Eq. (\ref{eq:fn}) is Friedmann 
equation. One should realize, however, that only the case 
$\rho\propto a^{-3}$, for pressureless matter, is possible.  

Let us now express Eq. (\ref{eq:fn}) in terms of the canonical 
variables. The Lagrangian of the system is
\begin{equation}
L=\frac{1}{2}\sum_{i}^{n}m_{i} \dot r_{i}^2
+G\sum_{i>j}\frac{m_im_j}{|r_{i}-r_{j}|}
+\frac{\Lambda}{6}\sum_{i}^{n} r_{i}^2.
\end{equation}
By assuming the cosmological principle this reads
\begin{eqnarray}
L&=&\frac{1}{2}A \dot S^2
+G \frac{B}{S}+\frac{\Lambda}{6}A S^2, \nonumber\\
 &=&\frac{A}{2\mu^2} \dot a^2
+G \frac{\mu B}{a}+\frac{\Lambda A}{6\mu^2} a^2. \label{eq:lc}
\end{eqnarray}
From this we can construct two phase spaces: one defined as $(S,P_{S})$
with $P_S=A\dot S$ and the other as $(a,P_{a})$ with
\begin{equation}
P_{a}=\frac{A}{\mu^2} \dot a \label{eq:momenton},
\end{equation}
which are equivalent to each other. In terms of the $(S, P_{S})$
variables, the Hamiltonian takes the form
\begin{equation}
H=E=\frac{ P_{S}^{2}}{2A}-\frac{GB}{S}-
\frac{\Lambda A S^2}{6};
\end{equation}
whereas Friedmann equation (\ref{eq:fn}) in terms of the canonical
variables $(a,P_a)$ reads 
\begin{equation}
\frac{P_{a}^{2}}{a^{2}}+
\left(\frac{A}{\mu^2}\right)^2 \left(\frac{k}{a^{2}}-\frac{\Lambda}{3}
-\frac{8\pi G}{3} \rho \right) =0. \label{eq:fc}
\end{equation}
In order to quantize this system we must write down the wave equation
\begin{equation}
\hat H\psi (S)= \left(\frac{\hat P_{S}^{2}}{2A}-\frac{GB}{S}-
\frac{\Lambda A S^2}{6}\right )\psi (S)=E\psi(S). \label{eq:sn}
\end{equation} 
This is equivalent to request physical states, $\psi(a)$, which
vanish within Friedmann equation (\ref{eq:fc}), i.e.
\begin{equation}
\left[\frac{1}{a^2}\hat P_{a}^{2}+
\left(\frac{A}{\mu^2}\right)^2 \left(\frac{k}{a^{2}}-\frac{\Lambda}{3}
-\frac{8\pi G}{3} \rho \right)\right]\psi(a)=0,
\label{eq:wdn}
\end{equation}
with $\hat P_{a}=-i\hbar{\partial}/{\partial a}$.
Solutions to this equation are the quantum states of the Newtonian
universe. Another proposal of quantum Newtonian cosmology can be
found in Ref. \cite{gibbonsc:gnus}.

\section{Quantum Cosmology in (3+1) Dimensions}
We describe now the differences in the quantum level between Newtonian
and relativistic cosmologies.
It is a known fact that Friedmann equation (\ref{eq:fn}) can also be 
obtained from Einstein equations for a space with cosmological
constant and the Robertson-Walker metric \cite{wald:gnus}.
For this metric, the kinetic term in the Lagrangian from 
general relativity is
\begin{equation}
L=-\frac{\alpha}{2} a\dot a^2.
\end{equation}
Therefore, for this case the canonical momentum is given by
\begin{equation}
\frac{\partial L}{\partial \dot a}=
\Pi_{a}= -\alpha a\dot a. 
\label{eq:momento}
\end{equation}
Note that this differs from the canonical momentum from classical
mechanics in Eq. (\ref{eq:momenton}). Now, by using Eq. (\ref{eq:momento})
Friedmann equation (\ref{eq:fn}) becomes
\begin{equation}
{\cal H}=\frac{\Pi_{a}^{2}}{a^{4}}+ \left(\frac{1}{\alpha}\right)^2 
\left(\frac{k}{a^{2}}-\frac{\Lambda}{3}
-\frac{8\pi G}{3} \rho \right) =0.\label{eq:fg}
\end{equation}
In the classical level, both Eqs. (\ref{eq:fg}) and (\ref{eq:fc}) 
describe the same dynamical system as just different auxiliary
variables were used to rewrite them. Though this is irrelevant
in the classical regime, it yields notable differences quantumly.

Within the canonical formalism Eq. (\ref{eq:fg}) represents a
first-class constraint \cite{D1:gnus}. According to that
formalism, the quantity $\cal H$ is substituted by an operator
and the physical states, $\psi(a)$, are found as the null vectors of 
this operator. Therefore, the equation determining the physical states is
\begin{equation}
\left[ \frac{\hat \Pi_{a}^{2}}{a^{4}}+ \left(\frac{1}{\alpha}\right)^2 
\left(\frac{k}{a^{2}}-\frac{\Lambda}{3}
-\frac{8\pi G}{3} \rho \right) \right]\psi(a)=0,\label{eq:wdg}
\end{equation}
with $\hat \Pi_{a}=-i\hbar{\partial }/{\partial a}$.
According to general relativity, this equation determines the quantum
state of the universe. This is the so-called Wheeler-De Witt
equation \cite{vilenki:gnus}.

As it can be seen, Eq. (\ref{eq:wdn}) differs from (\ref{eq:wdg}).
This is because the classical phase space of both systems is
different. Thus, in the classical level, Newtonian and relativistic
in (3+1) dimensions cosmologies are equivalent, but quantumly they are 
different things. As an example, by considering the $\rho=0$ and 
$\Lambda\not=0$ case we can see that: for the Newtonian cosmology
the system behaves as an oscillator both in the classical and quantum levels. 
However, for the relativistic cosmology the system in the classical level 
behaves as an oscillator, but in the quantum level it is no longer
an oscillator. It could be thought that the inequivalence arises from 
the canonical momentum definition and that, by using another quantization
formalism, the equivalence could be preserved. However, due to the difference
in actions, after carrying out the path integrals one would obtain that 
difference again.

\section{Cosmology in (2+1) Dimensions}
Now we concentrate in the connection between Newtonian cosmology
and cosmology in a (2+1) dimensional space. Let us then consider
the metric
\begin{equation}
ds^{2}= dt^{2}-a^{2}\left(\frac{dr^{2} }{1-kr^{2}}+r^2 d\theta^2\right).
\label{eq:metrica}
\end{equation}
For pressureless matter, and this metric, Einstein equations are
\begin{eqnarray}
\left(\frac{\dot a}{a}\right)^2&=&
-\frac{k}{a^2}+\Lambda+8\pi G \rho,\\
\,\,\,\,\frac{\ddot a}{a}&=&\Lambda.
\end{eqnarray}
By using Eq. (\ref{eq:metrica}) the kinetic term in the Lagrangian from 
general relativity becomes
\begin{equation}
L=- \frac{\beta}{2} \dot a^{2};
\end{equation}
so that the canonical momentum is
\begin{equation}
P_a=- \beta \dot a. 
\end{equation}
From this, Friedmann equation becomes
\begin{equation}
\frac{P_{a}^{2}}{a^{2}}+
\left(\frac{1}{\beta} \right)^2 \left(\frac{k}{a^{2}}-
\Lambda-8\pi G \rho\right) =0. \label{eq:fc2}
\end{equation}
By quantizing this system we find physical states, $\psi(a)$,
such that
\begin{equation}
\left[\frac{\hat P_{a}^{2} }{a^2}+
\left(\frac{1}{\beta}\right)^2 \left(\frac{k}{a^{2}}-\Lambda
-8\pi G \rho \right)\right]\psi(a)=0.
\label{eq:wdn2}
\end{equation}
Clearly, up to ordering terms, Eq. (\ref{eq:fc}) is analogous to
(\ref{eq:fc2}). Nevertheless, in two dimensions, for pressureless
matter one has $\rho\propto a^{-2}$, which is different to the
three-dimensional case; but if we take $\rho\approx 0$ and perform
the changes
\begin{equation}
\beta^{-1} \to \frac{A}{\mu^2},\quad 
\Lambda \to \frac{\Lambda}{3},
\end{equation}
then Eq. (\ref{eq:fc2}) becomes equal to (\ref{eq:fc}) and Eq. 
(\ref{eq:wdn2}) analogous to (\ref{eq:wdn}). That is, for the matterless 
case, both in the classical and quantum levels Newtonian cosmology can 
be mapped into the relativistic one in (2+1) dimensions.

\section{Wave Function of the Newtonian Universe}
Let us now look at the wave function of the Newtonian universe.
Because $S$ and $a$ are positive, the wave equation is defined
only in the positive axis. This leads one to require appropriate 
boundary conditions. By imposing $\psi(\infty)=0$, one must also 
impose $\psi(0)=0$ \cite{Landau:gnus}.

We first consider the case of matter dominated by a negative 
cosmological constant; i.e. $\Lambda=-|\Lambda|$ and $\rho\approx 0$. 
For this, Eq. (\ref{eq:sn}) reads
\begin{equation}
\hat H\psi(S)= \left [\frac{-\hbar^{2}}{2A}\frac{\partial^{2}}
{\partial S^{2}}-\frac{A\Lambda}{6}S^{2}\right]\psi(S)=E\psi(S).
\label{eq:ecs1}
\end{equation}
By introducing the variable
$z=\left (\frac{A}{\hbar}\sqrt{\frac{|\Lambda|}{3}}\,\,\right)^{1/2}\!\!S$,
Eq. (\ref{eq:ecs1}) can be rewritten as
\begin{equation}
\left [\frac{\partial^{2}}{\partial z^{2}}+
\left (\frac{2E}{\hbar\sqrt{\frac{|\Lambda|}{3}}}-z^{2}\right)\right]\psi(z)=0;
\label{eq:ecs2}
\end{equation}
which has solutions
\begin{equation}
\psi_{n}(z)=H_{2n+1}(z)e^{-z^{2}/2},\quad n=0,1,\dots
\end{equation}
where $H_N(z)$ is the Hermite polynomial of order $N$; and energy spectrum
\begin{equation}
E_{n}=\hbar\sqrt{\frac{|\Lambda|}{3}}\left(2n+\frac{3}{2}\right).
\end{equation}
Note that this does not depend on $A$.

Now we turn to the dust case, i.e. $\rho=\rho_{0}/a^3$ and $\Lambda=0$.
For this, the wave equation reads
\begin{equation}
\left[\frac{\partial^{2}}{\partial S^2}+\left(\frac{2EA}{\hbar^2}+
\frac{2ABG}{\hbar^2}\frac {1}{S} \right) \right]\psi(S)=0. 
\end{equation}
By using $z=\left(\sqrt{\frac{-8EA}{\hbar^2}}\,\,\right)\!S$, one gets to
the equation 
\begin{equation}
\left[z\frac{\partial^{2}}{\partial z^2}+
\left(\gamma-\frac{z}{4}\right)\right]\psi(z)=0, 
\end{equation}
where $\gamma=\frac{2ABG}{\hbar^{2}}\sqrt{\frac{\hbar^2}{-8EA}}$. 
In this case the complete set of solutions vanishing at the
origin and infinity is
\begin{equation}
\psi_{n}(z)=e^{-z/2}zL^{1}_{n}(z);
\end{equation}
where $L^{1}_{n}(z)$ is the associate Laguerre polynomial of order $n$;
whereas the energy spectrum is given by
\begin{equation}
E_{n}(z)=-\frac{AB^2G^2}{2\hbar^2 (n+1)^2}.  
\end{equation}
We can see then that there are two cases where exact solutions to the 
wave equation of the Newtonian cosmology can be obtained. For the same 
two cases in the relativistic cosmology in (3+1) dimensions no exact
solutions are known.

It is remarkable that, for a cosmology with cosmological constant and
pressureless matter, there are theories which in the classical regime
yield the same dynamics, but in the quantum level are inequivalent.
This phenomenon has been also found in other physical systems 
\cite{pi:gnus}.

\section{Conclusions}
It was shown that despite, for pressureless matter, Newtonian cosmology 
is equivalent to a relativistic one in (3+1) dimensions, in the quantum 
level they are inequivalent. It was also shown, however, that especial 
cases exist where the quantum Newtonian cosmology can be mapped to a 
quantum cosmology in (2+1) dimensions. At the end, two cases were presented 
where the exact wave function of the Newtonian universe could be found.

\section*{Acknowledgments}
The authors acknowledge the ICN-UNAM for its kind hospitality and AZ also 
for financial support.

\end{document}